\pdfoutput=1
\documentclass[aps,prd,a4paper,11pt,notitlepage,superscriptaddress,longbibliography,nofootinbib]{revtex4-1}
\usepackage{microtype}
\usepackage{graphicx}
\graphicspath{{./figures/}}

\usepackage{mathtools}
\usepackage{amssymb}
\DeclareMathOperator{\sgn}{sgn}
\DeclarePairedDelimiter{\abs}{\lvert}{\rvert}

\usepackage[dvipsnames]{xcolor}
\usepackage[utf8]{inputenc}
\usepackage[T1]{fontenc}
\usepackage[pdfusetitle,bookmarksopen,colorlinks,allcolors=blue]{hyperref}

\begin{document}
\title{Kink-antikink collisions in hyper-massive models}

% LTeX: enabled=false
\author{F.~M.~Hahne}
\email{fernandomhahne@gmail.com}
\affiliation{Programa de Pós-Graduação em Física, Universidade Estadual de Santa Cruz,\\
Campus Soane Nazaré de Andrade, 45662-900, Ilhéus, Brazil}

\author{P.~Klimas}
\email{pawel.klimas@ufsc.br}
\affiliation{Departamento de Física, Universidade Federal de Santa Catarina,\\
Campus Trindade, 88040-900, Florianópolis, Brazil}
% LTeX: enabled=true

\begin{abstract}
    We study topological kinks and their interactions in a family of scalar field models with a double well potential parametrized by the mass of small perturbations around the vacua, ranging from the mass of the $\phi^4$ Klein-Gordon model all the way to the limit of infinite mass, which is identified with a non-analytic potential.
    In particular, we look at the problem of kink-antikink collisions and analyze the windows of capture and escape of the soliton pair as a function of collision velocity and model mass.
    We observe a disappearance of the capture cases for intermediary masses between the $\phi^4$ and non-analytic cases.
    The main features of the kink-antikink scattering are reproduced in a collective coordinates model, including the disappearance of the capture cases.
\end{abstract}

\maketitle

\section{Introduction}

Classical solutions of scalar fields are interesting due to their applications in a wide range of physical phenomena, from condensed matter systems to large cosmological settings.
Topological defect solutions, also known as topological solitons, are specially significant due to their applications and intricate dynamics~\cite{Manton:2004tk}.
In the $1+1$ dimensional case, the sine-Gordon model and the $\phi^4$ Klein-Gordon model are classical examples of theories supporting topological defects known as kinks.
The interaction of topological kinks in those and other models has been an active topic of research for over four decades, with many fascinating phenomena still to be explained~\cite{Kevrekidis:2019zon}.
The sine-Gordon model is remarkable for being an integrable model, allowing its kinks to collide elastically.
Meanwhile, the $\phi^4$ model has inelastic collisions featuring resonances cases, where a kink-antikink pair annihilates, oscillates a few times, and then escapes.
Recently, the resonance windows, among other features of kink-antikink interaction, have been better understood using a moduli space approach~\cite{Manton:2021ipk,Manton:2020onl,Pereira:2020jmi,Pereira:2021gvf,Adam:2021gat,Adam:2023qgx,Blaschke:2023mxj}.

Topological defects, like the kinks in the $\phi^4$ theory, are usually infinite in size, i.e.\ the field reaches the vacuum values only asymptotically through an exponential tail $~e^{-mr}$, where $m$ is the field mass and $r$ is the distance.
A fascinating exception is the case of compactons: solutions contained in a compact support, reaching the vacuum at a finite distance.
Compactons were first discovered for a modified Korteweg–De~Vries (KdV) equation~\cite{Rosenau:1993zz}, but have since also been found in relativistic models with non-differentiable potential at its minima.

Potentials that are non-differentiable at the minima, also known as non-analytic or V-shaped potentials, were first discussed in the continuum limit of mechanical models where the motion is limited by rigid barriers~\cite{Arodz:2002yt}.
Since then, it was shown that in certain cases the first Bogomol'nyi-Prasad-Sommerfield (BPS) submodel of the Skyrme model~\cite{Adam:2017pdh}, is equivalent to a scalar field theory with a non-analytic periodic potential~\cite{Klimas:2018woi}.
The collision of compact kinks in such model was recently studied~\cite{Hahne:2023dic}.
It was shown that, at least for that model, the compact kink-antikink scattering does not have resonance cases and that the transition from the cases of capture to the cases of escape of the soliton pair has a fractal nature.
In the case of compactons, the distinct mathematical nature of those solutions constrains the construction of the kink-antikink moduli space, making the determination of internal models that may contribute to the scattering moduli space more nuanced.

In this paper, we approach the problem of topological compacton interactions from a different perspective, building on a previous result that a non-analytic potential with compact solutions can be obtained by continually deforming the $\phi^4$ potential function~\cite{Bazeia:2014hja}.
The parameter of the deformation is related to the mass of small perturbations around the model vacua in such a way that the non-analytic model corresponds to the limit of infinite mass.
The identification of non-analytic potentials with infinitely massive fields has also been made in the case of a single well potential~\cite{Arodz:2007jh}.
In this sense, we can say that non-analytic models correspond to the hyper-massive limit of more usual models.
By studying how the features of kink-antikink scattering change during the transition from the $\phi^4$ to a non-analytic model, we seek to deepen the understanding of the unique features of compactons and their interactions.

In this paper, we improve on our previous results on compacton scattering by using kink internal modes derived from first principles instead of phenomenological arguments.
Furthermore, the model in the present paper has a more traditional double well potential, while the potential in our previous works was periodic.
This allows for a more objective verification of which results are actually due to non-analytic nature of the potential.
We also show that the transition from the $\phi^4$ model to non-analytic models is not as straightforward as it appears.
In some senses, we found that the interaction of kinks in the $\phi^4$ and in the hyper-massive model have more in common with one another than with intermediary cases.

This paper is organized as follows.
In section~\ref{sec:model} we present the model studied, reviewing some of its properties and the properties of the kink solution.
We also solve for the first kink internal mode for all values of the potential parameter.
In section~\ref{sec:single-kink-moduli-space} we discuss the moduli space of a single kink, including its Derrick mode and its similarity with the first kink internal mode.
We present simulation results for kink-antikink collisions in section~\ref{sec:kink-antikink-collisions}, with special attention to the identification of cases of capture or escape of the soliton pair.
Some simulation results are explained through the moduli space approach in section~\ref{sec:kink-antikink-moduli-space}.
We summarize our conclusions in section~\ref{sec:conclusions}.

\section{Model}
\label{sec:model}

We consider a model for a scalar field $\phi$ in $1+1$ dimensions defined by the Lagrangian density
\begin{equation*}
    \mathcal{L} = \frac{1}{2} \partial_\mu \phi \, \partial^\mu \phi - V_\alpha(\phi)
\end{equation*}
where $V_\alpha(\phi)$ is a double well potential with minima at $\phi = \pm 1$.
The potential is the one defined in ref.~\cite{Bazeia:2014hja}, given explicitly by
\begin{align*}
    V_\alpha(\phi) = \frac{\sqrt{1 + \alpha(2+\alpha)(\phi^2-1)^2} - 1}{2\alpha},
\end{align*}
where the parameter $\alpha$ is a real positive number.
The limiting cases for the parameter $\alpha$ correspond to the potential of the $\phi^4$ Klein-Gordon model and an analogous non-analytic model:
\begin{align*}
    V_0(\phi) &\equiv \lim_{\alpha\to 0} V_\alpha(\phi) = \frac{1}{2} (\phi^2 - 1)^2, \\
    V_\infty(\phi) &\equiv \lim_{\alpha\to\infty} V_\alpha(\phi) = \frac{1}{2} \abs{\phi^2 - 1}.
\end{align*}
For simplicity, every time we talk about the cases $\alpha\to0$ and $\alpha\to\infty$ we leave implied the procedure of taking the limit and simply say that $\alpha=0$ or $\alpha=\infty$.

\begin{figure}
    \includegraphics{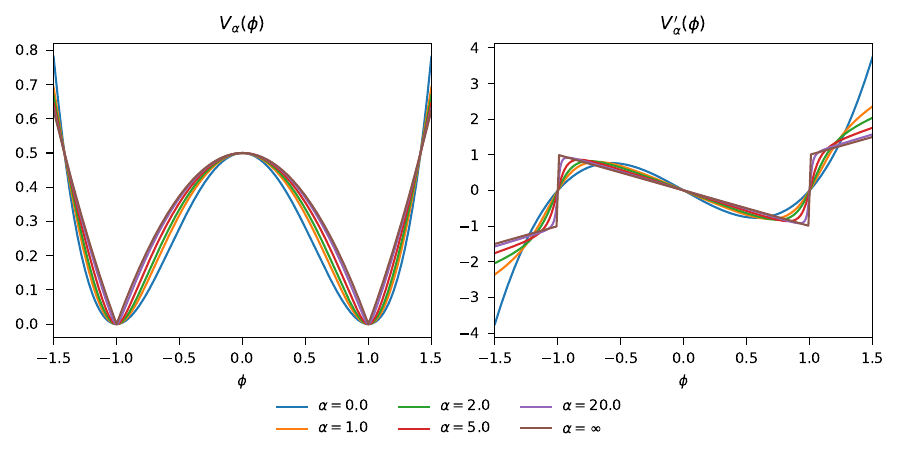}
    \caption{Potential $V_\alpha(\phi)$ (left) and its derivative $V'_\alpha(\phi)$ (right) for different values of $\alpha$.}
    \label{fig:double_well_potential}
\end{figure}

The field dynamics is determined by the Euler-Lagrange equation
\begin{equation*}
    \partial^\mu \partial_\mu \phi + V'_\alpha(\phi) = 0
\end{equation*}
where $V'_\alpha(\phi)$ is the potential derivative with the expression
\begin{equation*}
    V'_\alpha(\phi) = \frac{(2 + \alpha) \phi (\phi^2-1)}{\sqrt{1 + \alpha(2+\alpha)(\phi^2-1)^2}}.
\end{equation*}
Its limiting cases are
\begin{align*}
    V'_0(\phi) &= 2 (\phi^2 - 1) \phi, \\
    V'_\infty(\phi) &= \sgn(\phi^2-1) \phi.
\end{align*}
Note that $V'_\infty(\phi)$ is discontinuous at the potential minima, which is a defining characteristic of non-analytic potentials, allowing for the existence of compact solutions.
Also, we must impose that $V'_\infty(\pm 1) = 0$ so that the vacuum configurations $\phi=\pm1$ are solutions to the Euler-Lagrange equations.
This also agrees with the definition of derivative in the distributional sense.
Examples of the potential and its derivative for different values of $\alpha$ are shown in figure~\ref{fig:double_well_potential}.

Small field perturbations around the two vacua have squared mass equal to the second derivative of the potential:
\begin{equation*}
    m^2 = V_\alpha''(\pm 1) = 4 + 2\alpha.
\end{equation*}
Therefore, we can alternatively parametrize the potential by the field mass:
\begin{equation*}
    V_m(\phi) = \frac{\sqrt{m^2 (m^2 - 4) (\phi^2 - 1)^2 + 4} - 2}{2(m^2 - 4)}.
\end{equation*}
The usual $\phi^4$ potential can be recovered by taking the limit $m \to 2$ while the non-analytic case can be recovered in the limit $m \to \infty$.
In this sense, we say that models with non-analytic potential are a hyper-massive limit.
Note that this mass is expressed in dimensionless units, and can not be rescaled away.
It is an unavoidable parameter of the potential, which influences its shape beyond simple scale transformations.

There is yet another useful parametrization using the variable
\begin{equation*}
    \delta = \frac{\alpha}{1 + \alpha} \in [0, 1].
\end{equation*}
The parameter $\delta$ is useful when exploring the transition to the non-analytic case.
Since the range of $\delta$ is limited the non-analytic case can be reached by taking the limit $\delta \to 1$ from the left, a case we refer simply as the $\delta = 1$ case.
The potential in terms of $\delta$ is:
\begin{equation*}
    V_\delta(\phi) = \frac{(1 - \delta)}{2 \delta } \left(\sqrt{1+\frac{(2-\delta) \delta  \left(\phi ^2-1\right)^2}{(1- \delta)^2}}-1\right).
\end{equation*}
The corresponding mass parameter $m^2$ takes the form $m^2  = 2 (2-\delta) / (1-\delta)$.

\begin{figure}
    \includegraphics{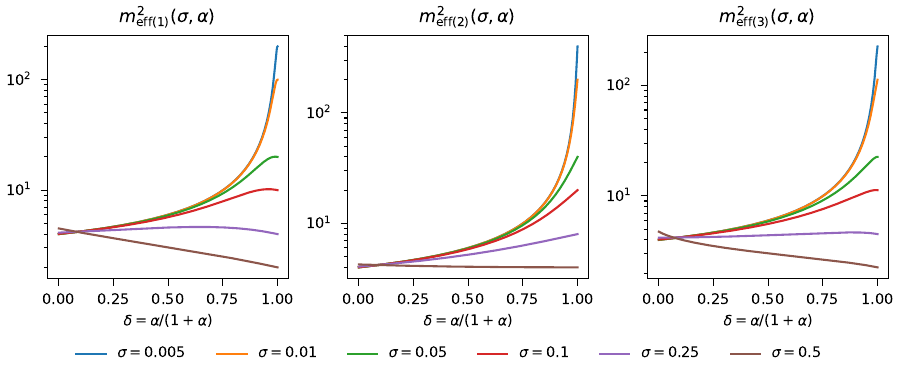}
    \caption{Effective mass as a function of the potential parameter for each weight function and different values of $\sigma$.}
    \label{fig:effective_mass}
\end{figure}

While the mass parameter $m^2$ becomes infinite in the limit $\alpha\to\infty$, it is worthy noticing that $m^2$ is determined by the potential second derivative precisely at the vacua.
Small field excitations oscillate around the vacua value, and therefore are also influenced by the potential for values slightly different.
To capture some of this influence, we can define an effective mass for oscillations around the vacua~\cite{Dorey:2023sjh}.
The effective mass for oscillations of amplitude $\sigma$ is defined as the integral
\begin{equation*}
    m^2_\text{eff}(\sigma, \alpha) = \int_{-\infty}^{\infty} d\phi\; w_\sigma(\phi) \, V''_{\alpha}(\phi)
\end{equation*}
where $w_{\sigma}(\phi)$ is a weight function centralized around the potential minimum $\phi=1$ and with parameter $\sigma$ controlling its width.
Since the potential is symmetric under $\phi \to -\phi$, the following also applies to minimum $\phi = -1$.
We consider three types of weight functions
\begin{align*}
    w_\sigma^{(1)}(\phi) &=\frac{1}{2\sigma}\theta\left(1-\frac{|\phi-1|}{\sigma}\right),\\
    w_\sigma^{(2)}(\phi) &=\frac{1}{\sigma}\left(1-\frac{|\phi-1|}{\sigma}\right)\theta\left(1-\frac{|\phi-1|}{\sigma}\right),\\
    w_\sigma^{(3)}(\phi) &=\frac{1}{\sigma\sqrt{\pi}}\exp\left(-\frac{(\phi-1)^2}{\sigma^2}\right),
\end{align*}
where $\theta$ is the Heaviside step function.
Each function is normalized to unity, i.e.\ $\int d\phi \,w_a(\phi)=1$.
Additionally, the first two functions have compact support $\phi\in[1-\sigma, 1+\sigma]$.
In figure~\ref{fig:effective_mass} we plot the $m^2_\text{eff}$ in dependence of the potential parameter $\delta=\alpha/(1+\alpha)$.
We see that for small enough $\sigma$, the effective mass is the largest for the non-analytic potential.
However, for larger values the behavior is not always monotonically increasing, although the maximum is close to $\delta=1$.
Therefore, for large amplitude oscillations, the hyper-massiveness is not the only relevant aspect of the potential.

Note that we have to use the generalized second derivative in the $\alpha=\infty$ case, which gives the following expression
\[
    V''_\infty(\phi)=-\sgn(1-\phi^2)+2\delta(\phi-1)+2\delta(\phi+1)
\]
where $\delta(\phi\pm1)$ are Dirac's delta functions.
In the hyper-massive case $\alpha=\infty$, the effective mass for each weight function is
\begin{align*}
    m^2_\text{eff(1)}(\sigma, \infty) &=2w_\sigma^{(1)}(1)=\frac{1}{\sigma},\\
    m^2_\text{eff(2)}(\sigma, \infty) &=2w_\sigma^{(2)}(1)=\frac{2}{\sigma},\\
    m^2_\text{eff(3)}(\sigma, \infty) &=\frac{2}{\sigma\sqrt{\pi}} \left[1+\exp\left(-\frac{4}{\sigma^2}\right)\right]+1-\operatorname{erf}\left(\frac{2}{\sigma}\right).
\end{align*}
The last case behaves as $m^2_\text{eff} \approx 2 w^{(3)}(1)=2 / (\sigma\sqrt{\pi})$ for $\sigma\ll1$.
Therefore, there is a finite effective mass gap unless $\sigma=0$.

\begin{figure}
    \includegraphics{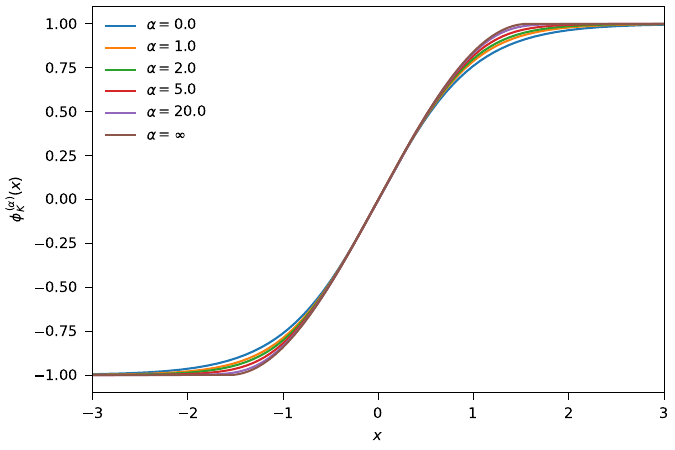}
    \caption{Kink profile $\phi^{(\alpha)}_K(x)$ for different values of $\alpha$.}
    \label{fig:kink}
\end{figure}

Since the potential is degenerate for all values of the parameter $\alpha$, topological kink solutions can be found by integrating the BPS equation
\begin{equation}
    \frac{d\phi_K^{(\alpha)}(x)}{dx} = \sqrt{2V_\alpha \big(\phi_K^{(\alpha)}(x)\big)} \label{eq:BPS}
\end{equation}
with appropriate boundary conditions.
For $\alpha\to0$ we get the usual $\phi^4$ kink
\begin{equation*}
    \phi^{(0)}_K(x) = \tanh x
\end{equation*}
while for $\alpha\to\infty$ we get the compact kink solution
\begin{equation*}
    \phi^{(\infty)}_K(x) =
    \begin{cases}
        -1 & \text{ if } x < -\frac{\pi}{2}, \\
        \sin x & \text{ if } -\frac{\pi}{2} \leq x \leq \frac{\pi}{2}, \\
        1 & \text{ if } x > \frac{\pi}{2}.
    \end{cases}
\end{equation*}
For other values of $\alpha$, the kink solution does not have an analytical expression, but it can be found by numerically integrating equation~\eqref{eq:BPS}.
Some examples are presented in figure~\ref{fig:kink}.
Due to the symmetry of the model, antikink solutions are simply $\phi^{(\alpha)}_K(-x) = -\phi^{(\alpha)}_K(x)$.
Note that the potential $V_{\infty}(\phi)$ for the region between the two minima, i.e.\ $\phi \in [-1,1]$, has the same shape that of the periodic potential of refs.~\cite{Klimas:2018woi,Hahne:2022wyl,Hahne:2023dic}.
Therefore, the compact kink has the same profile shape as the one in those works.

Kink profiles are usually formed by a core, skin, and tails~\cite{Karpisek:2024zdj}, with sizes in the target space estimated by the expressions
\begin{equation*}
    L_\text{core} = \frac{- 2 V''(\phi_\text{max})}{\abs{V'''(\phi_\text{max})}}, \quad
    L_\text{tail} = \frac{2 m^2}{\abs{V'''(\phi_\text{min})}}, \qquad
    L_\text{skin} = \sqrt{\frac{2 V'(\phi_\text{inf})}{\abs{V'''(\phi_\text{inf})}}}
\end{equation*}
where $\phi_\text{max}$, $\phi_\text{min}$ and $\phi_\text{inf}$ denote the maximum, minimum, and inflection points of the potential, respectively.
% The perturbative mass parameter $m^2$ is defined as the second derivative (curvature) of the potential evaluated at its minimum point. % Already defined before
The application of these formulas must be handled with caution.
While these formulas are effective for analytic potentials, they encounter difficulties with non-analytic case due to the non-existence of certain derivatives in the classical sense.
For example, as $\delta$ approaches $1$, $m^2$ approaches infinity according to the formula $m^2=\frac{2 (2-\delta )}{1-\delta}$. The formula $V'''(\phi_\text{min})=\frac{6 (2-\delta )}{1-\delta }$ reveals that the third derivative of the potential at the minimum is proportional to the same expression as the second derivative.
This implies that the tail length remains constant $L_\text{tail}=\frac{2}{3}$ for all values of $\delta$ except for $\delta=1$, where a compact kink without tails emerges.
The compact kink is entirely composed of a core region.

In order to compute $L_\text{core}$ we evaluate the potential derivatives at the maximum point, $\phi_\text{max}=0$.
We find that $V''(\phi_\text{max})=\delta-2$ and $V'''(\phi_\text{max})=0$, independently of  $\delta$.
Therefore, we must estimate the core size using instead the fourth order derivative $V^{(4)}(\phi_\text{max})=6(2-\delta)(1-\delta)^2$.
This leads to the expression for $L_\text{core}$:
\[
    L_\text{core}=\sqrt{\frac{-6V''(\phi_\text{max})}{V^{(4)}(\phi_\text{max})}}=\frac{1}{1-\delta}.
\]
This expression exhibits proper behavior for small values of $\delta$, but diverges as $\delta$ approaches $1$.
Although the parameter $L_\text{core}$ indicates a growing core region in the kink as the potential becomes non-analytic, its numerical values should not be taken at face value for $\delta>\frac{1}{2}$, where $L_\text{core}$ surpasses $2$.

The inflection points are determined by exact but lengthy expressions, which we will omit for brevity.
The right inflection point $\phi_\text{inf}$ begins at $\frac{1}{\sqrt{3}}$ when $\delta=0$ and approaches $1$ as $\delta$ tends towards $1$.
The skin size $ L_\text{skin}(\delta)$ is a decreasing function with values ranging from $L_\text{skin}(0)=\frac{\sqrt{2}}{3}$ to $L_\text{skin}(1)=0$.
This accurately reflects the disappearance of the skin region in the non-analytic limit of the potential.

These kink and antikink solutions are static, however moving kink solutions $\psi^{(\alpha)}_K(t, x, v)$ with velocity $v$ can be found by exploring the Lorentz symmetry of the model, such that a moving kink solution $\psi^{(\alpha)}_K(t, x, v)$ is given by
\begin{equation*}
    \psi^{(\alpha)}_K(t, x, v) = \phi^{(\alpha)}_K(\gamma(x - vt))
\end{equation*}
where $\gamma = (1-v^2)^{-1/2}$.

Another possibility for dynamical behavior is to have the kink internal modes to be excited.
We can find the kink internal modes by writing the field as a kink profile with a small oscillating perturbation in the form
\begin{equation}
    \phi(t, x) = \phi_K(x) + \cos(\omega t) \, \chi(x)
    \label{eq:perturbation}
\end{equation}
where $\omega$ is the internal mode angular frequency,
and we temporally stopped writing the dependency in $\alpha$ for simplicity.
Plugging this expression on the field equation
\begin{equation*}
    \partial_t^2 \phi(t, x) - \partial_x^2 \phi(t, x) + V_{\alpha}'\big(\phi(t, x)\big) = 0
\end{equation*}
and assuming the perturbation is small, we obtain
\begin{equation*}
    \underbrace{\partial_t^2 \phi_K(t, x) - \partial_x^2 \phi_K(t, x) + V'_{\alpha}\big(\phi_K(t, x)\big)}_{0}
    + \cos(\omega t) \left[ -\omega^2 - \frac{d^2}{dx^2} + V''_{\alpha}\big(\phi_K(x)\big) \right] \chi(x) = 0
\end{equation*}
which leads to the following Schrödinger-like equation for the perturbation
\begin{equation}
    \left[- \frac{d^2}{dx^2} + V''_{\alpha}\big(\phi_K(x)\big) \right] \chi(x) = \omega^2 \chi(x).
    \label{eq:schrodinger}
\end{equation}

In general, the first eigenvalue $\omega_0^2$ is zero and corresponds to the translational mode of the kink.
The second eigenvalue gives the frequency square of the first internal mode.
For the case $\alpha=0$, this is the well-known shape mode with frequency $\omega_1 = \sqrt{3}$.
The shape mode of the $\phi^4$ kink is very widely studied, therefore we will not discuss it in detail.

\begin{figure}
    \includegraphics{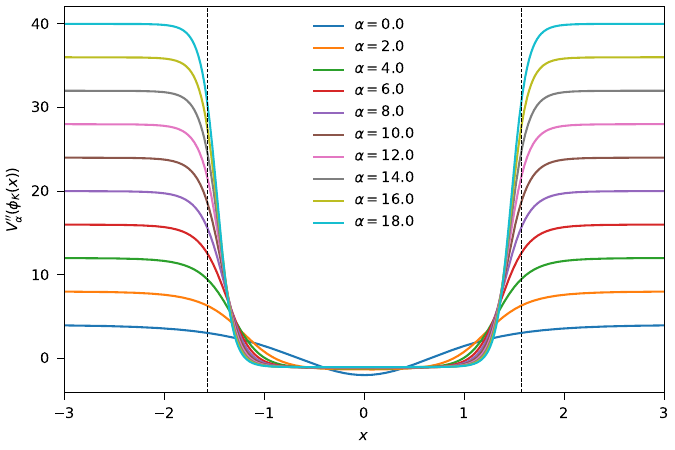}
    \caption{Function $V''_{\alpha}(\phi_K(x))$ which acts as potential in the Schrödinger type equation for small perturbations $\chi(x)$ of a kink profile. The vertical dashed lines indicate $x=\pm \pi/2$, which are the boundaries of the infinite well case for $\alpha\to\infty$.}
    \label{fig:schrodinger_potential}
\end{figure}

We can also get closed form expressions for the internal modes of the non-analytic potential in the hyper-massive limit.
The function $V''_{\alpha}(\phi_K(x))$ acts as the potential function for this equation.
For the non-analytic potential
\begin{equation*}
    V'_\infty(\phi) = \sgn(\phi^2 - 1) \phi.
\end{equation*}
This function has divergent derivative for $\phi = \pm1$,
therefore $V''_{\alpha}\big( \phi_K(x) \big)$ is infinite for $x = \pm \pi / 2$, because when $\alpha\to\infty$ the kink has a compact support and reaches the vacua $\phi=\pm1$ at these points.
This can also be visualized by looking the potential function for increasingly large values of $\alpha$, as in figure~\ref{fig:schrodinger_potential}.

Therefore, for the compact kink internal modes we must solve equation~\eqref{eq:schrodinger} for
\begin{equation*}
    V''_\infty(\phi^{(\infty)}_K(x)) =
    \begin{cases}
        -1 & \text{ if} \quad \mathopen{}-\frac{\pi}{2} < x < \frac{\pi}{2},\\
        \infty &\text{ otherwise,}
    \end{cases}
\end{equation*}
which is equivalent to the quantum mechanical problem of finding the wave function of a particle trapped in an infinite well.
It has been shown~\cite{Bazeia:2014hja} that the eigenvalues are $\omega_n^2 = n(n+2)$ for $n = 0, 1, 2, \ldots$,
which agrees with the textbook result for an infinite potential well.
Surprisingly, the first mode has frequency $\omega_1 = \sqrt{3}$ equal to the frequency of the shape mode in the $\phi^4$ model.
The eigenfunctions corresponding to each $\omega_n^2$ are simply trigonometric functions truncated to the kink support
\begin{equation*}
    \chi^{(\infty)}_n(x) =
    \begin{cases}
        \cos((n+1)x) &\text{ if } n \text{ is even and } -\frac{\pi}{2} < x < \frac{\pi}{2}, \\
        \sin((n+1)x) &\text{ if } n \text{ is odd and } -\frac{\pi}{2} < x < \frac{\pi}{2}, \\
        0 &\text{ otherwise.}
    \end{cases}
\end{equation*}
Note that $\chi^{(\infty)}_0(x) = d\phi^{(\infty)}_K(x) / dx$ is simply the translational mode.

By regularizing the potential we gain a simpler approach to analyzing the translational mode.
While directly perturbing the compact kink, as shown in equation~\eqref{eq:perturbation}, within the non-analytic model presents technical difficulties,
such as dividing the problem into separate spatial regions, leading to patched formulas, and handling Dirac deltas function arising from the second derivative $V''_\infty(\phi)$ of the potential,
the regularized approach offers a clearer path.
Here, the emergence of a square infinite well potential becomes evident, providing a well-defined method for obtaining the perturbation modes, including the translational mode.
Also note that the compacton internal mode  in this work is different from the phenomenological mode used in refs.~\cite{Hahne:2022wyl,Hahne:2023dic}, which was actually $\chi^{(\infty)}_1 + \frac{1}{2}\chi^{(\infty)}_3$.

\begin{figure}
    \includegraphics{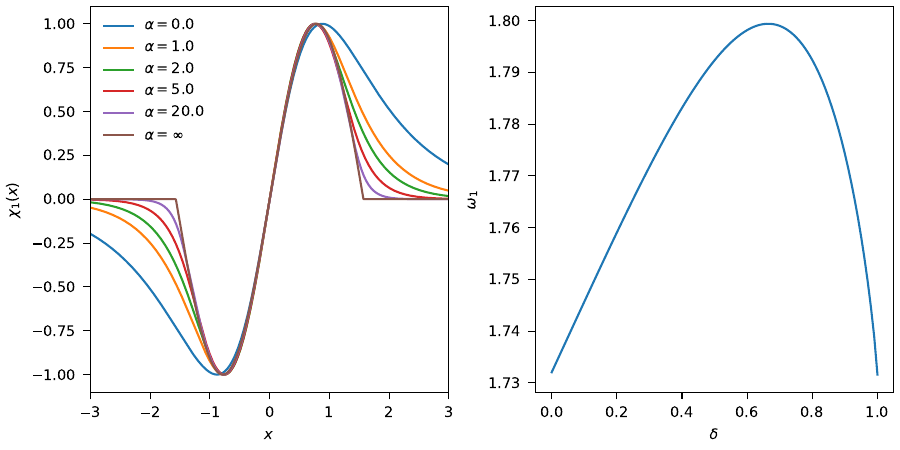}
    \caption{First kink internal model for selected values of $\alpha$ (left) and its frequency $\omega_1$ as function of $\delta=\alpha/(1+\alpha)$  (right). The mode has been normalized so that its peak has the unit value for all $\alpha$.}
    \label{fig:first_mode}
\end{figure}

For non-zero finite values of $\alpha$, the eigenvalue problem can not be solved analytically.
However, we can solve it numerically by discretizing the differential operator.
We present the first mode and its frequency for different $\alpha$ in figure~\ref{fig:first_mode}.
Much like the kink itself, the internal mode becomes compact in the hyper-massive limit $\alpha\to\infty$.
Note that, while the $\phi^4$ kink has a single internal mode, the increasing of the walls in
$V''_{\alpha}(\phi_K(x))$ allows for larger numbers of modes as we go towards the hyper-massive limit.
In fact, the compact kink has an infinite number of internal modes.
The dependency of the number of modes on $\alpha$ has been calculated in ref.~\cite{Bazeia:2014hja}.

\section{Single kink moduli space}
\label{sec:single-kink-moduli-space}

The moduli space approach, also known as collective coordinates approximation, is an important tool in the description of kink collisions since its earliest uses in the description of kink-antikink collisions in the $\phi^4$ model~\cite{Sugiyama:1979mi}.
In this approach, one writes a tentative expression for the field as a function of a finite number of degrees of freedom $q(t) = (q^1(t), q^2(t), \ldots)$ known as collective coordinates or moduli.
The time dependence of the field is then completely determined by $q(t)$, i.e.\ $\phi(t, x) = \phi(x; q(t))$.
The Lagrangian density for the field can be integrated to obtain a classical mechanics Lagrangian for the coordinates $q(t)$ with expression
\begin{equation*}
    L = \frac{1}{2} g_{ij}(q) \dot{q}^i \dot{q}^j - U(q)
\end{equation*}
in terms of an effective metric
\begin{equation}
    g_{ij}(q) = \int dx \, \frac{\partial \phi(x; q)}{\partial q^i} \frac{\partial \phi(x; q)}{\partial q^j} \label{eq:metric}
\end{equation}
and effective potential
\begin{equation}
    U(q) = \int dx \, \left[ \frac{1}{2} \left(\frac{\partial\phi(x; q)}{\partial x} \right)^2 + V(\phi(x; q)) \right] \label{eq:potential}
\end{equation}
in the space of the coordinates $q$.
Note that the metric is for the moduli space and must not be confused with the Minkowski spacetime metric.
The time evolution of $q(t)$ can be obtained by solving the Euler-Lagrange equations
\begin{equation}
    \ddot{q}^i + \Gamma^{i}_{jk}(q) \, \dot{q}^j \dot{q}^k + g^{il}(q) \frac{\partial U(q)}{\partial q^l} = 0 \label{eq:geodesic}
\end{equation}
where $g^{il}(q)$ is the inverse of the metric tensor and $\Gamma^l_{jk}$ are the Christoffel symbols given by
\begin{equation*}
    \Gamma^i_{jk}(q) = \frac{1}{2} g^{il}(q) \left[ \frac{\partial g_{lj}(q)}{\partial q^k} + \frac{\partial g_{lk}(q)}{\partial q^j} - \frac{\partial g_{jk}(q)}{\partial q^l} \right].
\end{equation*}

For a single kink configuration, we can explore the translational symmetry of the model to get a displaced soliton centered at $x=a$ with the expression $\phi^{(\alpha)}_K(x - a)$, which is a solution of the field equations provided that $a$ is a constant.
If we promote the kink position $a$ to a time-dependent degree of freedom, we obtain a non-relativistic moduli space with a single coordinate $q^1(t) = a(t)$.
It can be shown~\cite{Manton:2020onl} that the Lagrangian for this coordinate is simply the one for a massive particle with constant potential:
\begin{equation*}
    L = \frac{1}{2} M^{(\alpha)} \dot{a}^2 - M^{(\alpha)}
\end{equation*}
where
\begin{equation*}
    M^{(\alpha)} = \int_{-\infty}^{\infty} dx \left[ \frac{d\phi^{(\alpha)}_K(x)}{dx} \right]^2
\end{equation*}
is the kink rest energy, which is also its mass because we are using natural units.

A relativistic moduli space can be achieved by introducing an extra coordinate $b(t)$ to account for the Lorentz contraction, so that the field has the expression
\begin{equation*}
    \phi(x; a, b) = \phi^{(\alpha)}_K\big( b \, (x - a) \big).
\end{equation*}
In this case, the Lagrangian can be shown~\cite{Adam:2021gat} to have the expression
\begin{equation*}
    L = \frac{1}{2} M^{(\alpha)} b \dot{a}^2 + \frac{1}{2} \frac{Q^{(\alpha)}}{b^3} \dot{b}^2 - \frac{1}{2} M^{(\alpha)} \left(b + \frac{1}{b} \right)
\end{equation*}
where
\begin{equation*}
    Q^{(\alpha)} =\int_{-\infty}^{\infty} dx \, x^2 \left[\frac{d\phi_K^{(\alpha)}(x)}{dx} \right]^2
\end{equation*}
is the static kink second energy moment.
The resulting Euler-Lagrange equations have a stationary solution $\dot{a} = v$, $ b = \gamma = (1-v^2)^{-1/2}$, which reproduces exactly the field equation solution of a moving kink.
In the case $\dot{a} = 0$, one can also find oscillatory solutions with frequency $\omega_D^2 = M^{(\alpha)} / Q^{(\alpha)}$.

The frequency $\omega_D$ is the frequency of the Derrick mode, related to the small changes in the field caused by scale transformations of the kink solution.
The profile of the Derrick mode can be obtained considering the perturbation on the kink profile by a small size deformation, i.e.~$b$ close to one.
In fact, for $a=0$ and $b = 1 +\varepsilon$, the field has the Taylor series
\begin{equation}
    \phi^{(\alpha)}_K(bx) = \phi^{(\alpha)}_K(x) + \sum_{n=1}^{\infty} x^n \frac{d^n \phi^{(\alpha)}_K(x)}{dx^n} \frac{\varepsilon^n}{n!}.
    \label{eq:derrick-series}
\end{equation}
We identify the coefficient of first term in the series as the Derrick mode
\begin{equation*}
    \chi_D (x) = x \, \frac{d\phi^{(\alpha)}_K (x)}{dx}.
\end{equation*}

\begin{figure}
    \includegraphics{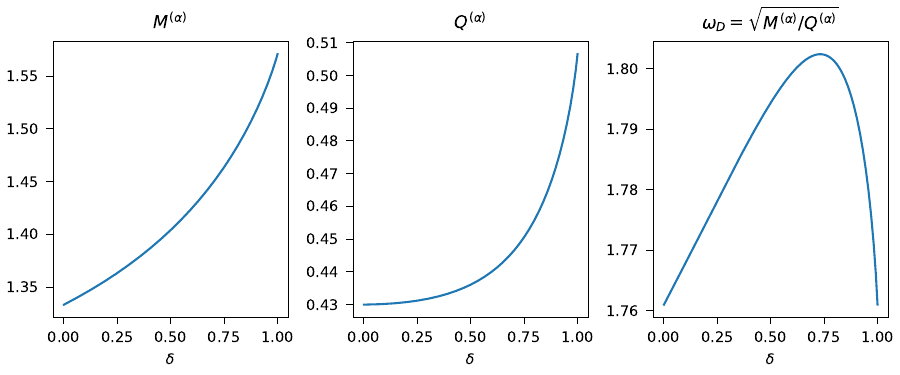}
    \caption{Kink mass $M^{(\alpha)}$, second energy moment $Q^{(\alpha)}$, and Derrick mode frequency $\omega_D$ as functions of $\delta = \alpha / (1 + \alpha)$.}
    \label{fig:derrick}
\end{figure}

In figure~\ref{fig:derrick} we present the dependency of the quantities $M^{(\alpha)}$, $Q^{(\alpha)}$, and $\omega_D$ on the parameter $\delta = \alpha / (1 + \alpha)$.
As the potential $V_\alpha(\phi)$ becomes steeper with larger $\alpha$ the kink narrows, leading to an increase in its gradient energy reflected on the kink mass $M^{(\alpha)}$.
The minimum mass, $M^{(0)}=4/3$ corresponds to the $\phi^4$ kink, while the maximum mass, $M^{(\infty)}=\pi/2$, is for the compact kink.
The antikink has a mass equal to the mass of the kink.
The minimum value of the second moment is $Q^{(0)}=(\pi^2-6)/{9}$, while the maximum value is $Q^{(\infty)}=\pi(\pi^2-6) / 24$.
We have
\begin{equation*}
    \frac{Q^{(\infty)}}{Q^{(0)}} = \frac{M^{(\infty)}}{M^{(0)}} = \frac{3\pi}{8}
\end{equation*}
causing the Derrick mode for the cases $\alpha=0$ and $\alpha=\infty$ to have identical frequencies,
\begin{equation*}
\omega_D=2\sqrt{\frac{3}{\pi^2-6}}\approx 1.7609,
\end{equation*}
similar to what was found for the frequency of the first internal mode, $\omega_1=\sqrt{3}\approx 1.7320$.

\begin{figure}
    \includegraphics{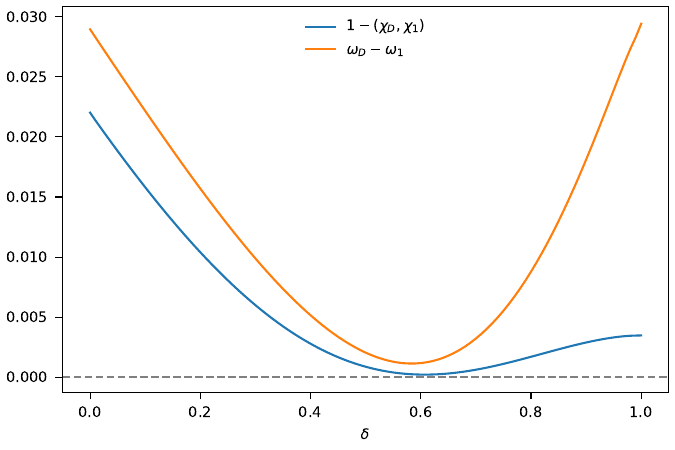}
    \caption{Quantities $1 - (\chi_D, \chi_1)$ and $\omega_D - \omega_1$ as functions of $\delta = \alpha / (1 + \alpha)$. These quantities are measures of the difference between the Derrick mode and the first internal mode.}
    \label{fig:derrick_vs_internal}
\end{figure}

In fact, the whole curve for $\omega_D$ as function of $\delta$ is strikingly similar to the dependency of the first mode frequency $\omega_1$ on $\delta$ (figure~\ref{fig:first_mode}).
It is known that for the $\phi^4$ model the kink shape mode is very similar to the Derrick mode, both in profile and frequency.
It is natural to inquire whether the observed similarity persists for all values of $\alpha$.
To quantify the similarity between the modes, we define the inner product
\begin{equation*}
    (\chi_D, \chi_1) = \frac{\int_{-\infty}^{\infty} dx \, \chi_D(x) \chi_1(x) }{ \left[\int_{-\infty}^{\infty} dx \, \left(\chi_D(x)\right)^2 \int_{-\infty}^{\infty} dx \, \left(\chi_1(x)\right)^2 \right]^{1/2} }
\end{equation*}
between the modes.
This inner product ranges from $0$ for entirely independent modes to $1$ for identical modes.
Therefore, $1 - (\chi_D, \chi_1)$ is a measure of the modes' independence, and so is the frequencies difference $\omega_D - \omega_1$.
We graph the dependency of the these quantities on $\delta$ in figure~\ref{fig:derrick_vs_internal}.
Although the two modes never become identical, they are very close for $\delta\approx 0.6$, which is also the region where $\omega_D$ and $\omega_1$ are maximum.
Once again, the differences in frequency are the same for $\alpha=0$ and $\alpha=\infty$, however, the inner product $(\chi_D, \chi_1)$ is closer to unity for the non-analytic model than in the $\phi^4$ model.

\section{Kink-antikink collisions}
\label{sec:kink-antikink-collisions}

\begin{figure}[h]
    \includegraphics{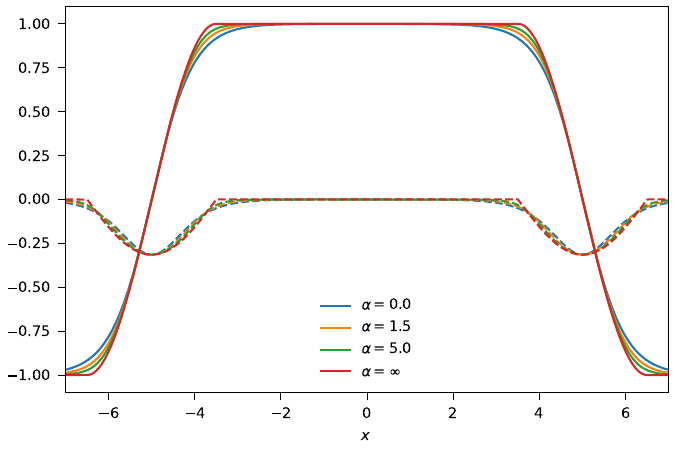}
    \caption{Initial conditions $\phi(0, x)$ (solid lines) and $\partial_t \phi(0, x)$ (dashed lines) of a kink-antikink collision for $v=0.3$, $x_0 = 5$, and some values of $\alpha$.}
    \label{fig:initial_conditions}
\end{figure}

The interaction between classical solutions offers important glimpses about the dynamics of non-linear models.
The most common scenario for studying interacting configurations in models with topological kinks is the collision between a kink and an antikink.
We study the problem of kink-antikink collisions in the center of momentum reference frame, where we can approximate the field initial condition by a superposition
\begin{align*}
    \phi(0, x) &= \psi^{(\alpha)}_K(0, x + x_0, v) - \psi^{(\alpha)}_K(0, x - x_0, -v) - 1,\\
    \partial_t\phi(0, x) &= \partial_t\psi^{(\alpha)}_K(0, x + x_0, v) - \partial_t\psi^{(\alpha)}_K(0, x - x_0, -v),
\end{align*}
provided that the separation $2x_0$ is large enough for the superposition to be a good approximate solution of the field equation, i.e.\ for the pair to be interacting weakly.
In the special case of compactons ($\alpha=\infty$), the functions $\psi^{(\alpha)}_K$ have support of size $\pi / \gamma$, so the superposition correspond an exact solution of the field equation when $x_0 \geq \pi / (2\gamma)$.
This initial condition contains a kink centered at $x = -x_0$ moving from the left to the right with velocity $v$, while an antikink centered at $x = x_0$ is moving from the right to the left with velocity of same absolute value, but opposite direction.
Some examples of the initial condition can be seen in figure~\ref{fig:initial_conditions}.

\begin{figure}[h]
    \includegraphics{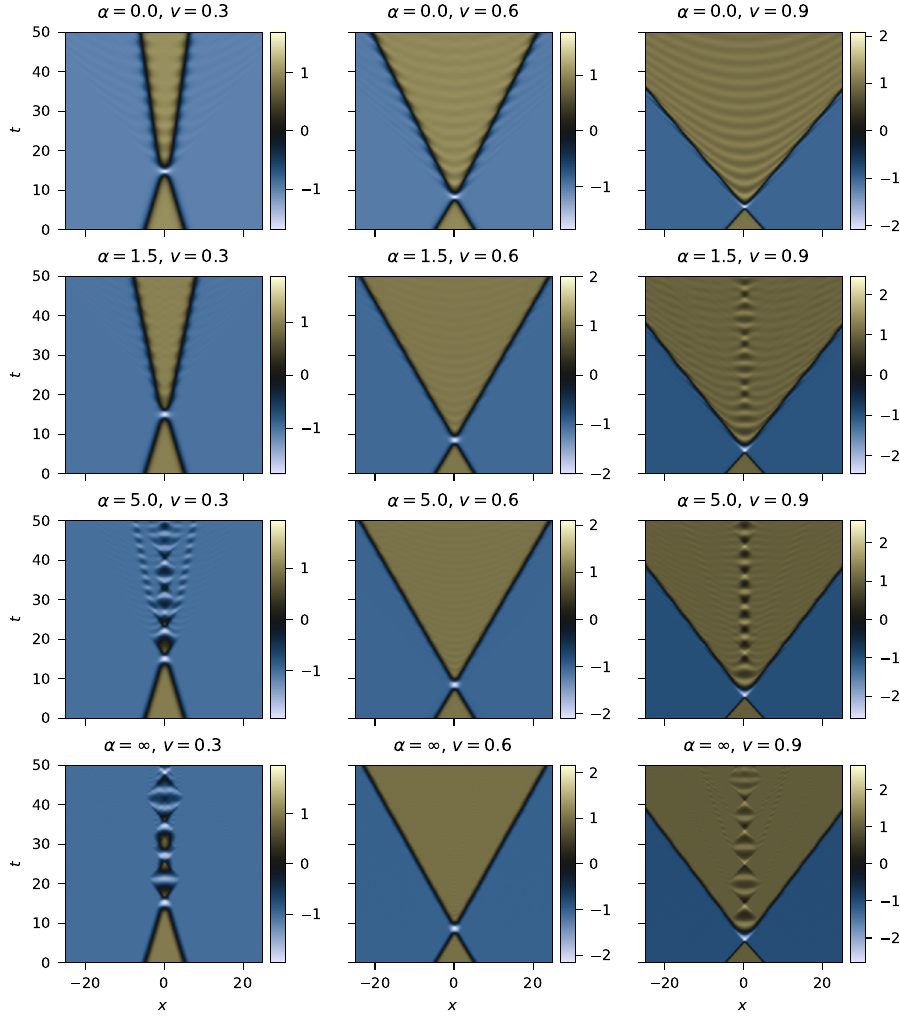}
    \caption{Simulation results for a kink-antikink collision with different values of $\alpha$ and $v$.}
    \label{fig:scattering_examples}
\end{figure}

Due to the non-linear and non-integrable character of the field equations, the dynamics of a kink-antikink interaction must be obtained by numerically evolving the initial conditions.
We discuss the numerical methods on the appendix.
We treat $\alpha$ and $v$ as parameters on which the simulation depends on, while fixing $x_0 = 5$.
Some examples of simulations for selected values of $\alpha$ and $v$ are presented as color maps on a spacetime diagram in figure~\ref{fig:scattering_examples}.
In general, all initial conditions initially evolve to the annihilation of the kink-antikink pair leading to a configuration where the field bounces between the vacuum values.
However, there are two broadly distinct outputs: either the field bounces indefinitely, a scenario we call \emph{capture}, or the field bounces one or more times and then a kink-antikink pair emerges from the configuration and separates, a scenario we call \emph{escape}.
In general, the capture case correspond to small velocities $v$, while the escape happens for larger values of $v$.

There is also emission of radiation, since the model is not integrable.
For small values of $\alpha$, the field behavior around the vacua is well approximated by the Klein-Gordon equation, which is a linear equation.
Therefore, the radiation for those cases resemble linear waves.
However, as the model becomes more massive, the field around the vacua is better approximated by the signum-Gordon equation, causing the radiation to be more localized.
In fact, the radiation spectrum of the signum-Gordon model is known to be dominated by compact oscillons~\cite{Hahne:2019ela}.

\begin{figure}[h]
    \includegraphics[width=\textwidth]{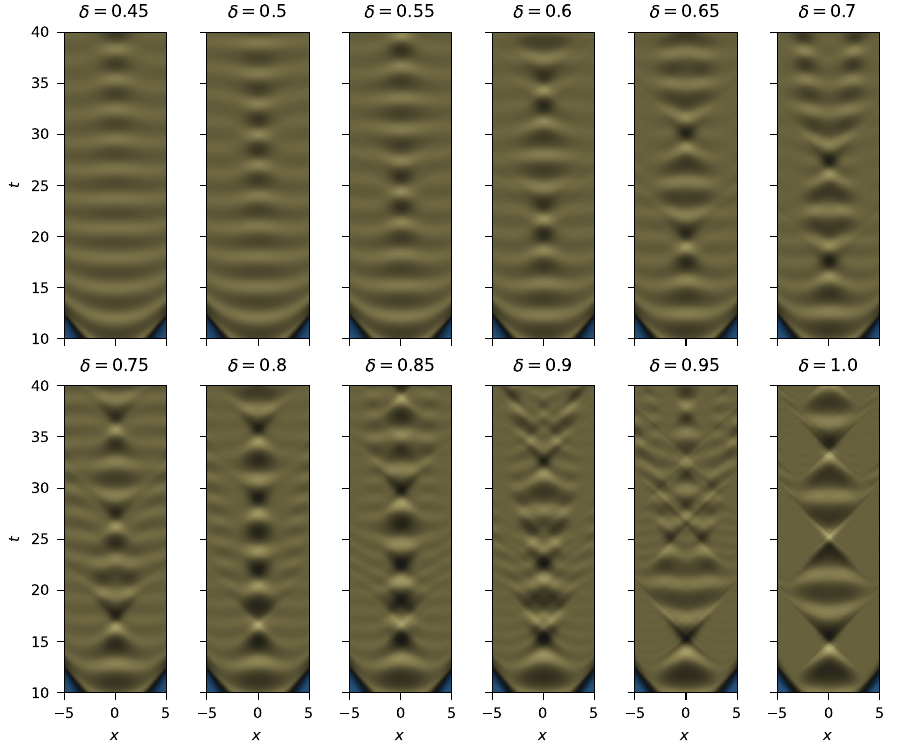}
    \caption{Central oscillon formed for kink-antikink collisions with $v=0.9$ and different $\delta=\alpha/(1+\alpha)$.}
    \label{fig:central_oscillon}
\end{figure}

In addition to the emission of radiation, we also observed the formation of a central oscillating compact structure for $\delta \gtrsim 0.5$ in the escape cases.
This result is monotonic, i.e.\ we found no windows of values of $\delta$ for which the central oscillon disappears for $\delta\gtrsim 0.5$, as can be seen in figure~\ref{fig:central_oscillon}.
This differs from the kink-antikink collision in a model with periodic potential, where the process becomes asymptotically elastic for large velocities, with compact shockwaves that decay into oscillons appearing for smaller velocities~\cite{Hahne:2023dic}.
The different dynamics for the escape leads to different collision byproducts, such as the central oscillon.
Also note, that the central oscillon for the non-analytic case $\delta=1$ has an oscillating support size, while the oscillons known analytically for the signum-Gordon model have support of constant size.
Similar results have been previously observed in the evolution of delta-like initial conditions, which can also lead to the creation of shockwaves~\cite{Hahne:2019odw}.
A unified framework for different kinds of compact oscillons and compact shockwaves, how they appear in the approach to the hyper-massive limit, and how they relate to configurations such as bounded kink-antikink pairs is a topic that needs further study, but lies outside the scope of this work.

\begin{figure}
    \includegraphics{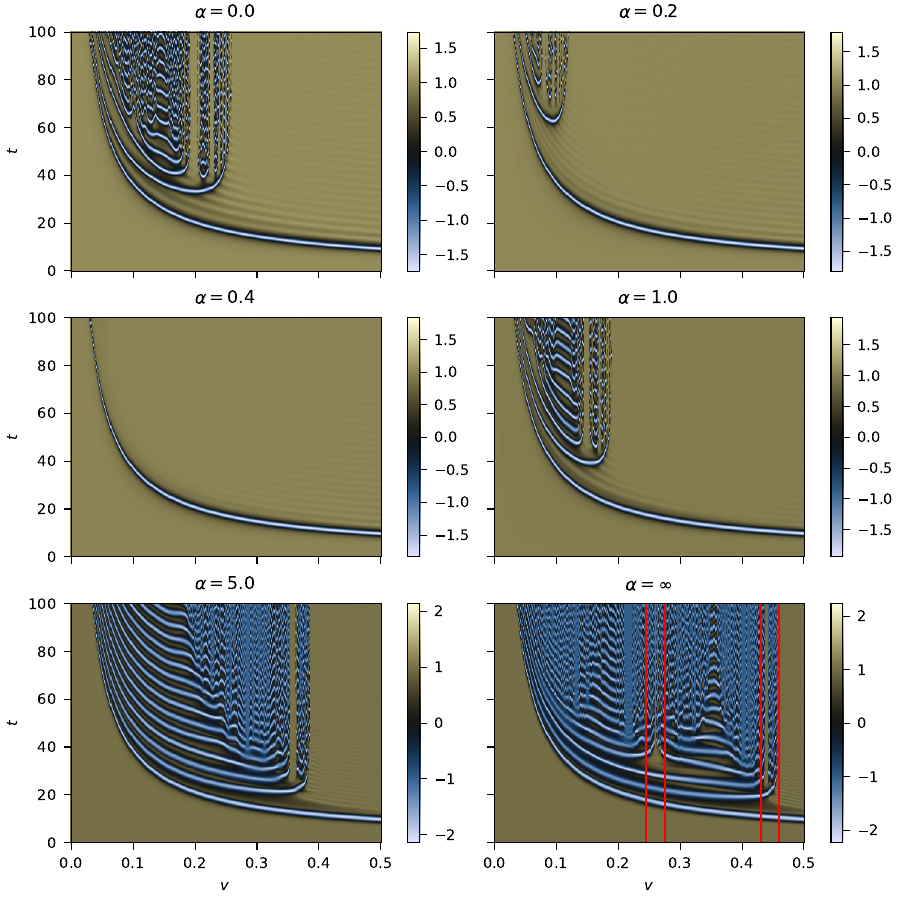}
    \caption{Simulation results for $\phi(t, 0)$ during a kink-antikink collision as functions of $v$ for selected values of $\alpha$. Regions delimited in red are shown in detail in figure~\ref{fig:scattering_middle_zoom}.}
    \label{fig:scattering_middle_examples}
\end{figure}

A more systematic classification of the scattering result as capture or escape can be accomplished without needing to analyze the full simulation results.
It is enough to concentrate on the field at the middle point between the kink and the antikink, which is the origin $x=0$.
From the function $\phi(t, 0)$ we can identify the capture cases as the ones in which the field at $x=0$ oscillates indefinitely, while the escape cases correspond to functions $\phi(t, 0)$ that remain close to the vacuum value $\phi=1$ after a few oscillations.
From our simulations we obtained the field at $x=0$ as a function of $t$ and $v$ for some values of $\alpha$, and present the results in figure~\ref{fig:scattering_middle_examples}.

\begin{figure}
    \includegraphics{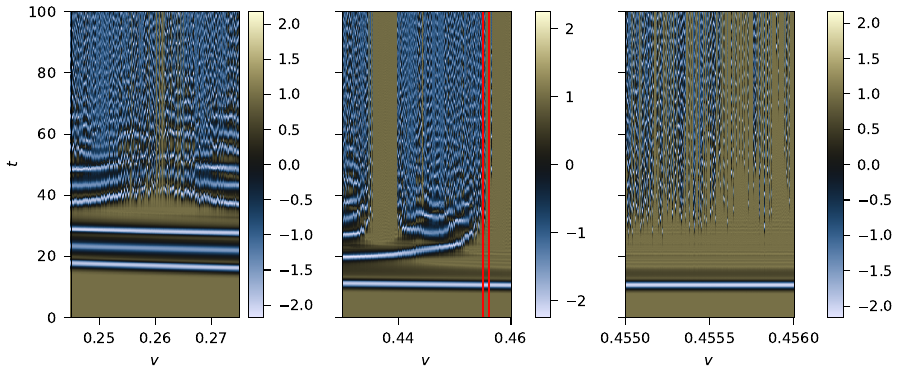}
    \caption{Field $\phi(t, 0)$ in the case $\alpha=\infty$. The first two panes are amplified images of the regions delimited in red in figure~\ref{fig:scattering_middle_examples}. The last panel is a further amplification into the region delimited in red in the second panel.}
    \label{fig:scattering_middle_zoom}
\end{figure}

For the case $\alpha=0$, we reproduce the picture of alternating windows of capture and escape of the $\phi^4$ model.
When we increase the value of $\alpha$ to $0.2$, we see that the range of velocities for which this pattern happens gets narrower.
For $\alpha=0.4$ ($m^2=4.8$), the pattern completely disappears, which means that for all velocities $v$ the kink-antikink pair annihilates and reemerges after a single bounce of the field, i.e.\ there are no capture cases.
If we keep increasing the value of $\alpha$, the capture cases reappear.
However, as we approach the hyper-massive limit, the pattern of escape windows becomes more straightforward.
In the non-analytic case $\alpha=\infty$, the capture and escape cases are almost neatly separated, with only two regions (delimited by red lines) with more than one bounce followed by escape, which we highlight in figure~\ref{fig:scattering_middle_zoom}.
However, the transition from the cases of capture to the cases of escape of the kink-antikink pair is not straightforward, and depends on the velocity in a rather subtle way.
In the case of compact kinks the transition from capture to escape happens in a fractal manner~\cite{Bazeia:2019tgt,Hahne:2023dic}.
In figure~\ref{fig:scattering_middle_zoom} we see that this fractal behavior is also present here, suggesting it is a general feature of compact kink-antikink collisions.

\begin{figure}
    \includegraphics{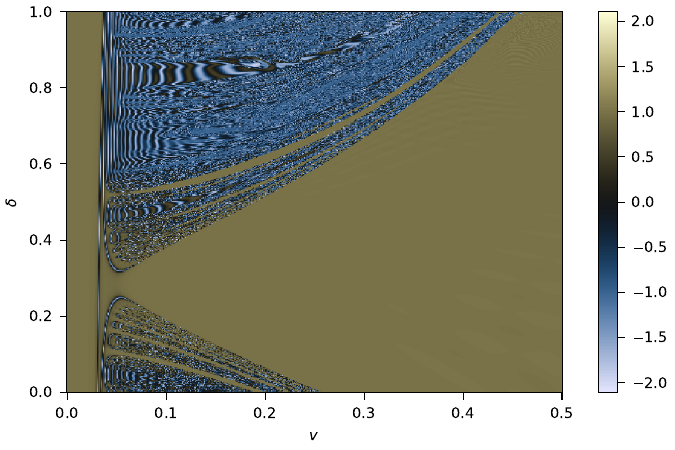}
    \caption{Kink-antikink scattering simulation results for $\phi(100, 0)$ as a function of $\delta = \alpha / (1 + \alpha)$ and $v$.}
    \label{fig:scattering_middle_final}
\end{figure}

At last, we take a more complete picture of the dependency on the potential parameter by looking at the field at the origin on a fixed ``final'' time, which we set as $t=100$.
When the kink-antikink pair escapes the field $\phi(100, 0)$ will have values close the vacuum value $\phi=1$.
However, for the capture cases, the field $\phi(100, 0)$ can have values very different due to its oscillatory behavior.
Therefore, the color map of $\phi(100, 0)$ as a function of $\alpha$ and $v$ can indicate the position of capture and escape cases in the parameter space $\delta \times v$.
In figure~\ref{fig:scattering_middle_final} we present the values of $\phi(100, 0)$ extracted from our simulations.

We observe that the capture windows get narrower until completely vanishing for $\delta$ close to $0.3$.
For more massive models, the capture windows reemerge in a pattern which is initially symmetric around the axis $\delta \approx 0.3$.
The symmetry stops holding approximately halfway towards the non-analytic case $\delta=1$.
The range of velocities $v$ where the capture cases are contained grows with the mass after its reemergence, and reaches its maximum for the non-analytic case.
We also see a band of escape cases (olive color in the plot) emerging between capture windows (blue) and moving to higher velocities as $\delta$ increases.
This escape cases are the ones seem in the middle panel of figure~\ref{fig:scattering_middle_zoom} for $\alpha=\infty$.

The disappearance and reemergence of the capture cases is a rather surprising feature of the transition towards the hyper-massive regime.
Since both the $\phi^4$ and the non-analytic model have similar results for the kink-antikink collisions, one could expect that the transition between them would be more straightforward.
The non-monotonic behavior of $\phi(100, 0)$ in relation to changes in $\delta$ indicates that some intricate mechanism happens exclusively for intermediary values of $\delta$.

It has been shown in ref.~\cite{Karpisek:2024zdj} that when a kink is core-less ($L_\text{core} = 0$), kink-antikink collisions are rather featureless processes, without resonance windows.
However, the lost of resonance windows happens monotonically as one reduces the core of the topological defects to zero.
Since we found a monotonic dependence of $L_\text{core}$, $L_\text{skin}$, and $L_\text{tail}$ for the family of potentials studied, the mechanism for the disappearance of capture cases is probably unrelated to the structure of the kinks considered here.

\section{Kink-antikink moduli space}
\label{sec:kink-antikink-moduli-space}

There is a lot of interest in finding effective descriptions of kink-antikink collisions without having to resort to full solutions of the field equations.
One approach that has found some success is the construction of kink-antikink moduli space in what is also known as the collective coordinate approximation.
Solving the dynamics for the collective coordinates is not necessarily easier than performing the full field simulations.
The evaluation of $g_{ij}(q)$ and $U(q)$ as well as the time evolution of the equations of motion for $q(t)$ are hard computational problems, subject to many nuanced numerical challenges, while the full field theory simulation is a more straightforward problem.
However, we still study collective coordinates approximations to the kink-antikink scattering because they allow us to more clearly identify which modes of the field are responsible for the phenomena observed in the full field simulations.
On the appendix we discuss more on the steps needed to numerically solve the collective coordinates models discussed below.

We first consider a moduli space where the field is described at all times by a simple superposition of kink and antikink, similar to the initial condition.
In this case, we have only one coordinate, the position $a$ of the antikink.
The configuration is symmetric around $x=0$, so the field is given by the formula
\begin{equation*}
    \phi^{(\alpha)}_{KAK}(x; a) = \phi^{(\alpha)}_{K}(x + a) - \phi^{(\alpha)}_{K}(x - a) - 1.
\end{equation*}
We call this the non-relativistic case since we do not take the Lorentz contraction into account.

\begin{figure}
    \includegraphics{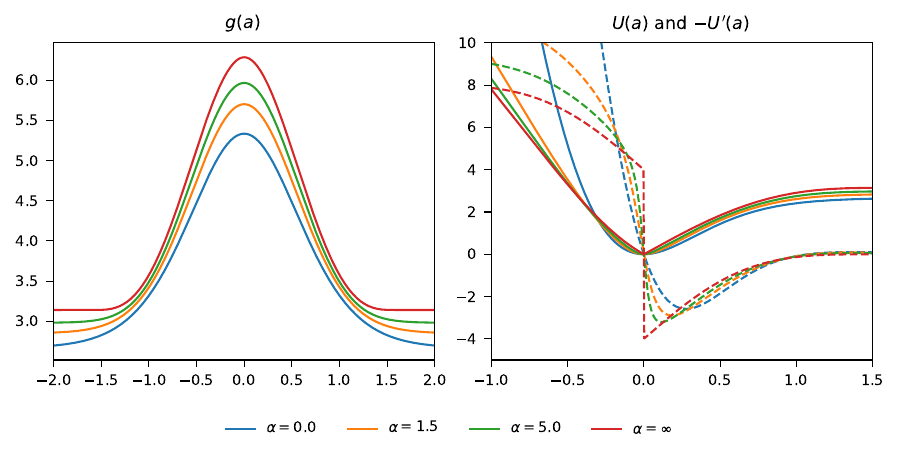}
    \caption{Metric $g(a)$ (left), potential $U(a)$ (right, solid line), and force $-U'(a)$ (right, dashed line) for collective coordinate $a$ for selected values of $\alpha$.}
    \label{fig:cc_metric_potential}
\end{figure}

Note that we only have closed form expression for $\phi^{(\alpha)}_{K}$ for the cases $\alpha=0$ and $\alpha=\infty$.
All the other cases have to be treated fully numerically.
Since the $\alpha=0$ case is the very widely studied $\phi^4$ model, we write the metric and potential as functions of $a$ for $\alpha=\infty$ only.
For the non-analytic model, we calculate the single metric component to be
\begin{equation*}
    g(a) \equiv g_{aa}(a) =
    \begin{cases}
        \pi + \sin(2\abs{a}) + (\pi - 2\abs{a})\cos(2\abs{a}) &\text{ if } \abs{a} < \frac{\pi}{2}, \\
        \pi & \text{ if } \abs{a} \geq \frac{\pi}{2},
    \end{cases}
\end{equation*}
while the potential is
\begin{equation*}
    U(a) =
    \begin{cases}
        -8a &\text{ if } a < -\frac{\pi}{2},\\
        -6a + 2\sin(2a) - (2a + \pi)\cos(2a) + \pi &\text{ if } -\frac{\pi}{2} \leq a < 0,\\
        2a + \sin(2a) & \text{ if } 0 \leq a < \frac{\pi}{2},\\
        \pi & \text{ if } a \geq \frac{\pi}{2}.
    \end{cases}
\end{equation*}
These are piecewise functions because they must take into account all the different possibilities for overlapping support of the kink solutions.
In figure~\ref{fig:cc_metric_potential} we present the graph of these functions, together with the numerical results for other values of $\alpha$.

For the non-relativistic model in this section, the position $a$ is the only degree of freedom, and therefore it carries all the configuration energy.
Since the effective mechanical system has conservation of energy, the motion of $a$ is always reflected by the potential $U(a)$, making so that all collisions result in escape of the kink-antikink pair.
However, the shape of the potential can still give us insight into the capture cases by imagining that the activation of unaccounted modes or radiation makes $a$ lose energy and get trapped at the potential well, oscillating around $a=0$.

\begin{figure}
    \includegraphics{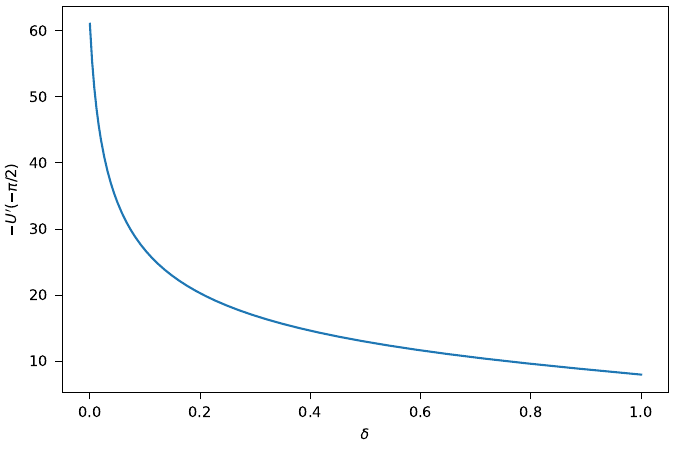}
    \caption{Force $-U'(-\pi/2)$ as a function of $\delta=\alpha/(1+\alpha)$.}
    \label{fig:cc_force}
\end{figure}

In particular, we can look at the force $-U'(a)$ to understand the higher critical velocity separating the capture from escape case in the non-analytic model in relation to models with smaller masses.
Since the force for negative $a$ is smaller for higher values of $\alpha$, there is less repulsion between the soliton pair while it bounces.
Figure~\ref{fig:cc_metric_potential} shows that the force grows as $a$ goes towards more negative values.
However, the slope of $-U'(a)$ is greater for smaller values of $\alpha$.
This makes the interaction more repulsive for small $\alpha$ when $a$ goes to more negative values.
Therefore, the soliton pair is more likely to be separated when $\alpha$ is small.
For the $\alpha=\infty$ case, the force $-U'(a)$ is constant for $a<-\pi/2$, having its maximum value of 8.
This is a consequence of the fact that compact kinks do not interact when their support do not overlap.
For other $\alpha$, the force keeps growing for more negative $a$ because the kinks interact even at large distances.
We compare the force for the whole range of $\delta = \alpha / (1 + \alpha)$ at $a = -\pi/2$ in figure~\ref{fig:cc_force}.
The interaction is indeed more repulsive for small $\delta$.
However, the behavior is monotonically decreasing in $\delta$, offering no explanation for the absence of capture cases for $\delta\approx0.3$.

We can improve on the description by adding more kink modes to the field expression.
A tentative expression for the field during a kink-antikink collision is a superposition with the first internal mode activated:
\begin{equation*}
    \phi^{(\alpha)}_{KAK}(x; a, c) = \phi^{(\alpha)}_K(x + a)- \phi^{(\alpha)}_K(x - a) - 1 + \frac{c}{f(a)} \left[\chi^{(\alpha)}_1(x + a) - \chi^{(\alpha)}_1(x - a)\right].
\end{equation*}
The collective coordinates are $a$ and $c$, which can be interpreted as the antikink center of momentum and internal mode amplitude, respectively.
The function is $f(a)$ is any function that behaves as $f(a) \sim a$ for small $a$, and as $f(a) \sim \pm 1$ for large $a$.
A function like this fixes the null-vector problem at $a=0$ while keeping the interpretation of $c$ as the amplitude of the internal mode when the solitons are separated~\cite{Manton:2020onl}.
Provided that the above properties are satisfied, the specific choice of $f(a)$ does not matter, since cases with different $f(a)$ can be mapped through a redefinition of the coordinate $c$.
A convenient choice is $f(a) = \phi^{(\alpha)}_K(a)$.

\begin{figure}
    \includegraphics{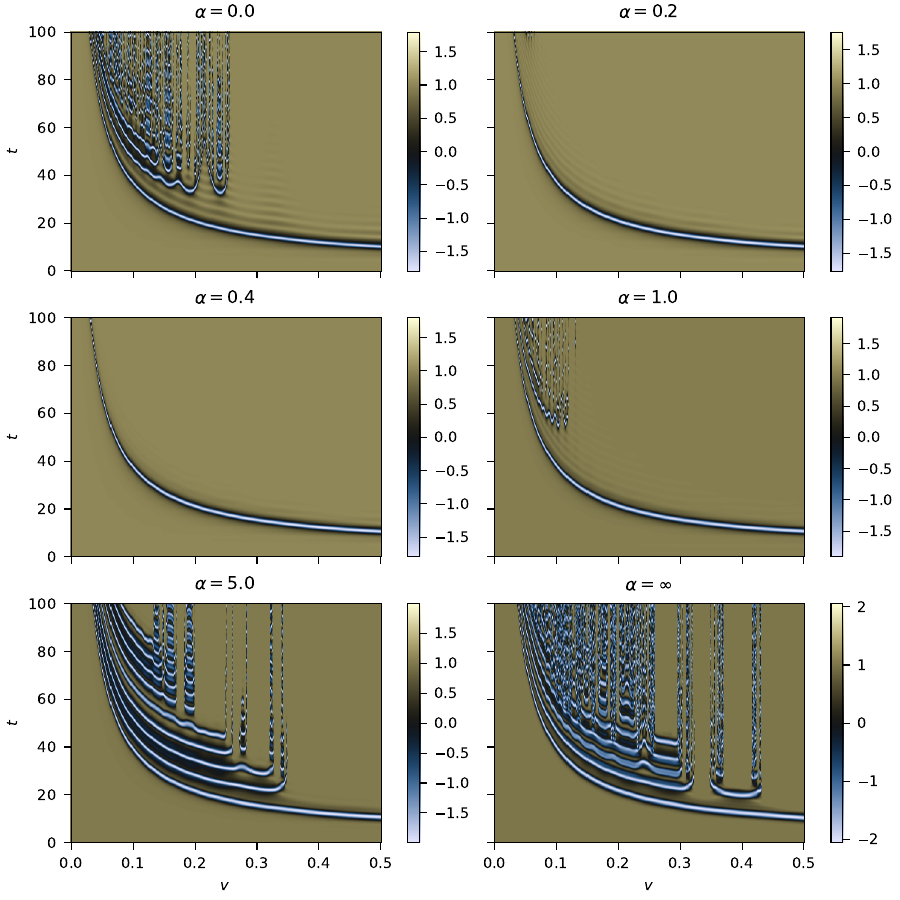}
    \caption{Collective coordinates predictions for $\phi(t, 0)$ as function of $v$ for selected values of $\alpha$.}
    \label{fig:cc_middle}
\end{figure}

The internal mode can store part of the collision energy, making so that the translational mode gets trapped in the potential well.
This translates into the presence of capture cases.
Similar to the analysis of the simulation results, we can identify scattering result as capture or escape by looking at the field at the origin $x=0$.
Using the numerical results for $a(t)$ and $c(t)$, we calculated $\phi(t, 0) = \phi_{KAK}(0; a(t), c(t))$ for different values of $\alpha$ and $v$.
In figure~\ref{fig:cc_middle} we present these results as function of time $t$ and scattering velocity $v$ for the same selected values of $\alpha$ used in figure~\ref{fig:scattering_middle_examples}.

The comparison between the simulation results in figure~\ref{fig:scattering_middle_examples} and the collective coordinates predictions in figure~\ref{fig:cc_middle} show reasonable agreement.
While the collective coordinates model studied here is too simple to be quantitatively predict all features of the field dynamics, it correctly predicted the disappearance and reemergence of capture windows observed in the simulations.
Since the only additional coordinate in relation to the previous model is the internal mode amplitude $c$, we can conclude that the first internal mode is the main driver of the dependence of the capture cases on $\alpha$ and $v$.

Another important find is that the collective coordinates approach works even though the internal mode makes the field derivative $\partial_x \phi(t, x)$ discontinuous at $x = \pm (\abs{a} + \pi/2)$ in the non-analytic case.
Even though the field equations demands $\partial_x \phi(t, x)$ to be continuous, the collective coordinates approach is more flexible because no second order derivative of $\phi$ appears in the metric or the potential.
Furthermore, the discontinuity of $\partial_x \phi(t, x)$ does not affect the continuity of $g_{ij}(q)$ and $U(q)$ due to the integration in the formulas~\eqref{eq:metric} and~\eqref{eq:potential}.

\begin{figure}
    \includegraphics{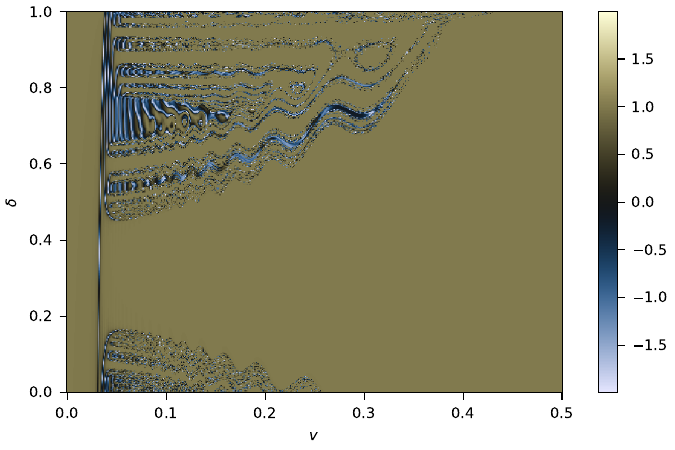}
    \caption{Field $\phi(100, 0)$ as predicted by the non-relativistic collective coordinates model with an internal mode.}
    \label{fig:cc_middle_final}
\end{figure}

We take a more complete look at the ultimate result of the collision predicted by the collective coordinates approximation in figure~\ref{fig:cc_middle_final}, where we present the field at $x=0$ and $t=100$ as a function of $\delta = \alpha/(1+\alpha)$ and $v$.
Comparing with the full simulation results in figure~\ref{fig:scattering_middle_final}, we see that the collective coordinates model indeed predicts the lack of capture cases around $\delta=0.3$, although it happens for a wider range of values of $\delta$.
We also see some escape windows happening for regions of the $v \times \delta$ parameter space where the kink-antikink pair should have been captured.
This indicates that there are even higher modes, or even radiation, contributing to the presence of annihilation cases.
However, the moduli space approach performs reasonably well even though it is a model with only two degrees of freedom.
We could attempt to improve our results for the cases with large $\alpha$ by including higher kink internal modes when they exist.
However, as discussed in the appendix, the time evolution in the moduli space is a very difficult computational problem that scales poorly with the number of coordinates.

The collective coordinate model above is non-relativistic in the sense it does not include Lorentz contraction.
Introducing a new coordinate $b(t)$ to account for the Lorentz contraction, one could propose that the field is once again the superposition
\begin{equation*}
    \phi^{(\alpha)}_{KAK}(x; a, b) = \phi_K^{(\alpha)} (b(x+a)) - \phi_K^{(\alpha)}(b(x-a)) - 1
\end{equation*}
However, this approach generates a null-vector problem because $\partial_b \phi^{(\alpha)}_{KAK}(x; 0, b) = 0$, which makes an entire row and column of the metric null, which in turn makes impossible to solve equation~\eqref{eq:geodesic}.

To sidestep this null-vector problem, it has been proposed that, instead of including the scale factor $b$, one writes $b=1+\varepsilon$ and consider the relativistic modes perturbatively~\cite{Adam:2021gat}, using a series expansion like the one in equation~\eqref{eq:derrick-series}.
However, we replace the coefficients $\epsilon^n / n!$ with independent mode amplitudes $B_n$, so that the field for a single kink is
\begin{equation*}
    \phi_K^{(\alpha)}(x - a) + \sum_n \frac{B_n}{n!} (x-a)^n \frac{d^n \phi_K^{(\alpha)}(x-a)}{dx^n}
\end{equation*}
and kink-antikink configuration can be modeled by the expression
\begin{equation*}
    \begin{split}
        \phi^{(\alpha)}_{KAK}(x; a, B_1, B_2, \ldots) &= \phi_K^{(\alpha)} (x+a) - \phi_K^{(\alpha)}(x-a) - 1 \\
        &\qquad + \frac{1}{f(a)} \sum_n \frac{B_n}{n!} \left[(x+a)^n \frac{d^n \phi_K^{(\alpha)}(x+a)}{dx^n} - (x-a)^n \frac{d^n \phi_K^{(\alpha)}(x-a)}{dx^n} \right]
    \end{split}
\end{equation*}
where each $B_n$ is a coordinate, and we once again introduced $f(a)$ to avoid a null vector problem.

\begin{figure}
    \includegraphics{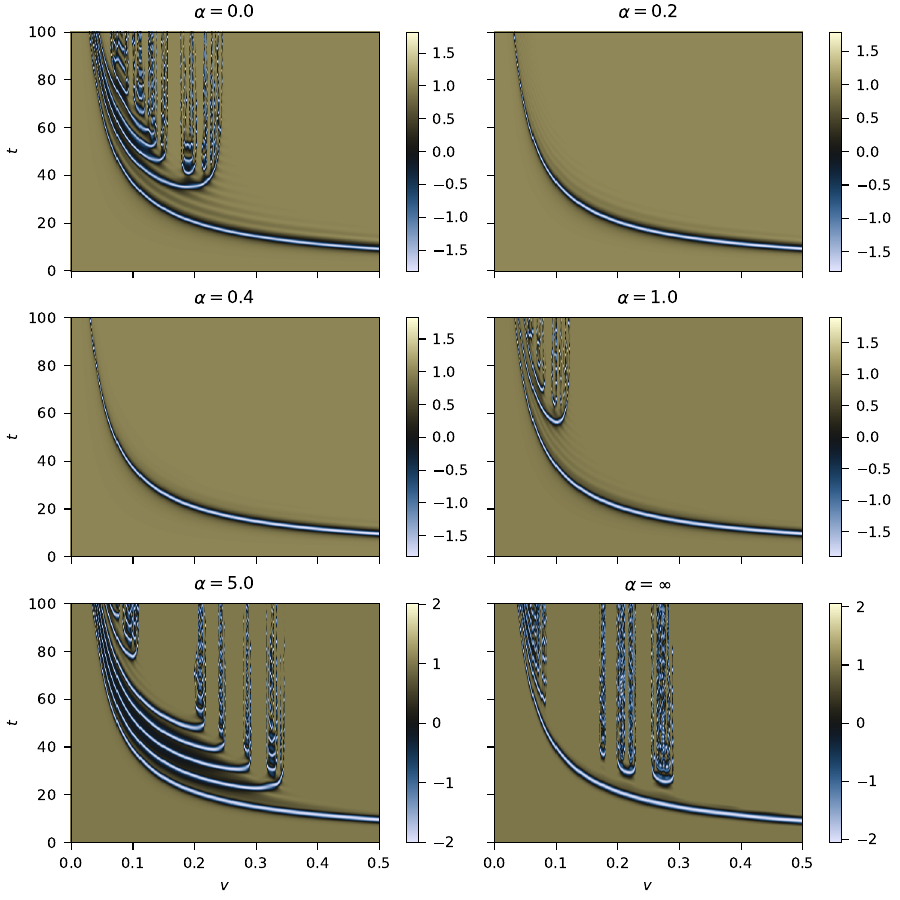}
    \caption{Field $\phi(t, 0)$ as predicted by perturbatively relativistic collective coordinate model with only the first Derrick mode.}
    \label{fig:cc_rel_middle}
\end{figure}

The collective coordinate model with only the first Derrick mode $B_1$ generates similar predictions to the model with the first internal mode, as seen in   figure~\ref{fig:cc_rel_middle}.
This is expected, because the Derrick mode is very similar to the first internal mode for all $\alpha$, as discussed in section~\ref{sec:single-kink-moduli-space}.

In some scalar field models, the collective coordinates predictions were significantly improved by the addition of higher modes.
This means, truncating the sum at some $n \geq 2$.
However, this approach is not possible in the hyper-massive model because $x^2 d\phi_K^{(\infty)} / dx$ is a discontinuous function.
Differently to the discontinuity in the first derivative of the internal mode, this discontinuity is in the mode itself, causing a divergence in the metric.
Note that, for small $a$
\begin{equation*}
    \lim_{a\to0} \frac{1}{f(a)} \left[ (x+a)^2 \frac{d^2 \phi_K^{(\alpha)}(x+a)}{dx^2} - (x-a)^2 \frac{d^2 \phi_K^{(\alpha)}(x-a)}{dx^2} \right]
    = 4x\frac{d^2 \phi_K^{(\alpha)}(x)}{dx^2} + 2x^2 \frac{d^3 \phi_K^{(\alpha)}(x)}{dx^3}.
\end{equation*}
The third derivative of the kink profile is proportional to a Dirac delta in the case $\alpha=\infty$.
This Dirac delta shows up square in the metric component $g_{B_2 B_2}$ for $a=0$, causing a divergence.
Therefore, the perturbative approach to relativistic moduli space is limited to the first mode in the hyper-massive regime.
Even for finite $\alpha$, as the mass increases, the metric becomes very large and causes numerical difficulties.
Therefore, in this work we do not explore coordinates beyond $B_1$.

\section{Conclusions}
\label{sec:conclusions}

In this paper we studied how the transition from the traditional $\phi^4$ model to a model with non-analytic potential influences static and interacting kinks.
We did this through a family of potentials parametrized by the mass of small perturbations around the vacua configurations, in a way that allows the non-analytic case to be identified as the infinite mass case.
In the non-analytic case, the kink and its internal modes have compact support, i.e.\ they are non-trivial only inside a compact region of space.
We observed that the internal and Derrick modes of the compact kink have identical frequencies to the case of the $\phi^4$ model.
It was also possible to generalize to the whole family of potentials the similarity between the first internal mode and the Derrick mode.

In the case of interaction kinks, we focused on the classification of kink-antikink collision results in escape or capture cases.
Our simulations showed that, as the mass increases, the window of velocities containing the capture cases get narrower until eventually disappearing.
However, if we continue to grow the mass, the capture cases reappear and their window of velocities grows, reaching its maximum value for the infinite mass case.
This rather surprising result suggests that the $\phi^4$ model and the non-analytic model have more in common with one another than with intermediary models.
An explanation for these similarities demands further investigation and is a possible follow up to this work.

To better understand the underlining causes of the capture cases, we analyzed the kink-antikink scattering through a moduli space approach.
We found that the inclusion of the first kink internal mode in the description is enough to explain qualitatively the dependence of the capture cases on the collision parameters, including the absence of capture when the potential parameter $\delta$ has values close to 0.3.
Similar results hold when replacing the internal mode with the first Derrick mode in a perturbatively relativistic approach.
However, we could do not expand the moduli space to include higher Derrick modes due to a discontinuity in such modes in the hyper-massive case.
This highlights how, even though the collective coordinate approximation is very useful even in non-analytic models, some techniques developed for more traditional models do not generalize.

\begin{acknowledgments}
    We thank R.\ Thibes for helpful comments and discussions.
    This study was financed in part by the Coordenação de Aperfeiçoamento de Pessoal de Nível Superior -- Brasil (CAPES) -- Finance Code 001 and by the Conselho Nacional de Desenvolvimento Científico e Tecnológico -- Brasil (CNPq).
\end{acknowledgments}

\appendix
\section{Numerical methods}

The numerical work was done using the Julia programming language~\cite{Bezanson:2014pyv} and the library DifferentialEquations.jl~\cite{Rackauckas:2017}.

For the full field equations, we discretize the spatial position $x$ with steps $\Delta x$ and approximate the derivatives using fourth-order finite differences.
The resulting coupled second order equation for the variables $\phi_n(t) = \phi(t, x_n)$ are evolved in time with a sixth-order Kahan-Li symplectic method~\cite{Kahan:1997} with time steps $\Delta t = 2 \Delta x / 5$.
We used spatial steps ranging from $\Delta x = 0.01$ to $0.05$.
Other methods and step sizes were tested and yielded consistent results.

In the collective coordinate approximation, we calculate the integrals for $g_{ij}(q)$ and $U(q)$ numerically.
Since the calculation of $g_{ij}(q)$ and $U(q)$ is computationally expensive, we compute it on a fixed grid for the collective coordinates and obtain the values from a cubic B-spline interpolation.
This approach is inspired by the one described in ref.~\cite{Adam:2022kla}.
With the metric and the potential given by such interpolations, we solved the equations of motion for $q$ using the velocity Verlet symplectic method~\cite{Verlet:1967}.
We tuned the numerical algorithms so that increases in the numerical precision (generally trough decreases in step size) did not significantly change the systematic results in figures~\ref{fig:cc_middle},~\ref{fig:cc_middle_final}, and~\ref{fig:cc_rel_middle}.
Therefore, the overall results represent a physically correct result, even though an independent implementation could find different trajectories for specific cases, leading to differences in the fine details.
This is due to the high susceptibility of the collective equations to small changes, since they represent very intricate dynamical systems.

Since the computation of $g_{ij}(q)$ and $U(q)$ becomes too expensive for high number of collective coordinates, we limited ourselves to moduli spaces with one or two dimensions by only including the first internal mode.
For example, generating figure~\ref{fig:cc_middle_final} with a resolution of $501\times 501$ data points required solving equation~\eqref{eq:geodesic} a total of 251001 times, as well as finding 501 kink solutions and their internal modes, which lead to 501 different metric and potentials functions, each one calculated from a 176000 point grid.
In other works~\cite{Adam:2021gat,Hahne:2023dic} the computation was done at every step of the time evolution, however this approach is also too computationally demanding for our needs, because we need to explore different values of $\alpha$, making the total number of scenarios to be considered way larger than in previous works.

\bibliography{bibliography}

\end{document}